# A Sub-Electron-Noise Multi-Channel Cryogenic Skipper-CCD Readout ASIC

Fabricio Alcalde Bessia*, Troy England*, Hongzhi Sun, Leandro Stefanazzi, Davide Braga,
Miguel Sofo Haro, Shaorui Li, Juan Estrada, Farah Fahim

*Abstract*—The *MIDNA* application specific integrated circuit (ASIC) is a skipper-CCD readout chip fabricated in a 65 nm LP-CMOS process that is capable of working at cryogenic temperatures. The chip integrates four front-end channels that process the skipper-CCD signal and performs differential averaging using a dual slope integration (DSI) circuit. Each readout channel contains a pre-amplifier, a DC restorer, and a dual-slope integrator with chopping capability. The integrator chopping is a key system design element in order to mitigate the effect of low-frequency noise produced by the integrator itself, and it is not often required with standard CCDs. Each channel consumes 4.5 mW of power, occupies 0.156 mm$^2$ area and has an input referred noise of 2.7 µV$_{rms}$. It is demonstrated experimentally to achieve sub-electron noise when coupled with a skipper-CCD by means of averaging samples of each pixel. Sub-electron noise is shown in three different acquisition approaches. The signal range is 6000 electrons. The readout system achieves $0.2$ e$^-$ RMS by averaging 1000 samples with MIDNA both at room temperature and at 180 Kelvin.

## I. INTRODUCTION

Skipper-Charge Coupled Devices (CCDs) are used for very sensitive applications, especially in particle physics due to their ultra-low noise performance. The non-destructive readout capability of skipper-CCDs has enabled a record low noise floor measurement of 0.068 $e^-_{rms}$ by averaging over 4000 reads [1]. A very good performance is routinely obtained when combining skipper-CCDs with the Low-Threshold Acquisition (LTA) controller [2]. For example, the Sub-Electron-Noise Skipper-CCD Experimental Instrument (SENSEI) [3] has been using skipper-CCDs with this readout system for searching for dark matter. Importantly, increasing the amount of sensitive mass by adding more skipper-CCDs can improve the detection probability of the experiment. Experiments like the Dark Matter in CCDs (DAMIC-M) [4], the Neutrino Interaction Observation with a Low Energy Threshold Array (νIOLETA) [5], and the Observatory of Skipper-CCDs Unveiling Recoiling Atoms (OSCURA) [6], [7] all plan to use hundreds or thousands of skipper-CCDs. In particular, the OSCURA

F. Alcalde Bessia and T. England contributed equally to this work.

F. Alcalde Bessia is with Instituto Balseiro and Instituto de Nanociencia y Nanotecnología INN (CNEA-CONICET), R8402AGP San Carlos de Bariloche, Argentina (e-mail: falcalde@ib.edu.ar).

T. England, H. Sun, L. Stefanazzi, D. Braga, S. Li, J. Estrada and F. Fahim are with the Fermi National Accelerator Laboratory, Pine & Kirk St, Batavia, IL 60510 (e-mail: tengland@fnal.gov).

M. Sofo Haro is with Universidad Nacional de Córdoba, Comisión Nacional de Energía Atómica (CNEA) and CONICET, X5000HUA Córdoba, Argentina (e-mail: miguelsofoharo@mi.unc.edu.ar).

experiment will use 24000 skipper-CCDs (28 Gpix) inside a low-temperature vessel. The LTA controller is based on high-speed discrete electronics, which would be prohibitively large and high power if scaled to read out thousands of video channels. The Multi-channel Integrating Dual-slope Noise-Aware (MIDNA) ASIC solves the problem of scalability by demonstrating sub-electron skipper-CCD readout at the chip level with cryogenic capability and unique system design.

This paper presents the MIDNA ASIC. MIDNA includes four signal processing channels, a digital controller, and several standalone blocks for testing purposes. The paper is organized as follows. Section II gives a detailed description of the MIDNA architecture and circuit implementation. A particular emphases is made on the noise performance and flexibility, showing the standard readout scheme and two interesting variations. Then, section III shows the experiments that were carried out using the MIDNA chip and demonstrates the noise reduction when applying the skipper technique along with the different readout schemes. Finally, section IV gives the conclusions of the work.

## II. DESIGN ARCHITECTURE

A popular signal processing channel used for reading out traditional scientific CCDs is based on dual slope integration (DSI) [8], [9], sometimes called *differential averaging*. DSI is one of many techniques to perform correlated double sampling (CDS) [10], [11], [12], [13]. With DSI the subtraction is computed in the analog domain with the advantage of optimal filtering of white noise versus measurement time and a full rejection of reset noise [14], [15]. The frequency response of a simple DSI system has been previously derived in [16]

$$|H_{DSI}(f)| = \frac{2A}{\pi T_S f} \sin^2(\pi T_S f) \quad (1)$$

where $A$ is the integration gain ratio and $T_S$ is the integration time of both the signal and the baseline levels. The total integration time is $T_T = 2 \times T_S$. The gain ratio is given by $A = T_S/RC$, where $R$ is the input resistor to the integrator and $C$ is the feedback resistor for the integrator.

At frequencies lower than $1/2\pi T_T$, the magnitude can be approximated utilizing the small angle approximation for sin.

$$|H_{LF}(f)| = 2\pi A T_S f \quad (2)$$

The result is a first-order high pass filter created by the CDS action. At frequencies higher than $1/2\pi T_T$, an approximation for the peaks of the magnitude is derived by setting the sin term to 1.

$$|H_{HF}(f)| = \frac{2A}{\pi T_S f}. \quad (3)$$

The result is a first-order low pass filter derived from the integrator. The frequency response of both approximations is based on the integration time. To reduce the noise bandwidth while keeping the gain constant, the integration time must be increased and integrator passives adjusted accordingly. The result will be the lowering of the center frequency. Eventually, the center frequency will reach the 1/f noise corner and the noise will no longer be reduced with longer integration times. Thus the lowest noise that has been achieved using this technique was approximately 2 e⁻ RMS [9], [17].

By using a skipper-CCD it is possible to overcome this issue. In a skipper-CCD, a given charge packet can be moved back and forth under the floating gate without destroying it and without losing charge. This means that several samples of the baseline and signal level can be taken and averaged, and each sample can be readout using a simple DSI circuit. The method—explained in detail in [16]—relies on averaging N equivalent samples, each one taken during a short period of time. The value of the center frequency of the readout noise bandwidth is maintained, avoiding the contribution of the $1/f$ noise. It has been demonstrated in [16] that by using this skipper readout technique, the RMS noise can be reduced a factor $1/\sqrt{N}$, with the theoretical reduction only limited by some practical aspects like losses in charge transfer and dark current noise.

*A. Large Signal Channel Concept*

Fig. 1 shows a simplified schematic view of one MIDNA signal processing channel. The channel is based on implementing and supporting DSI. The main blocks of each channel are the preamplifier, DC restorer, buffer, and dual-slope integrator. Fig. 2 shows a transient simulation of a MIDNA channel working through the operational phases. First the integrator is held in reset by Sreset, as named in Fig. 1, and the DC restorer switch, S1, is closed until the baseline voltage from the CCD is settled at the output of the preamplifier. Next, the negative integration begins. Sreset and S1 open. S3 and S5 close to connect the output of the buffer to the inverting input of the integrator and the reference voltage to the non-inverting input of the integrator. After the negative integration is complete, the charge collected by the CCD pixel is moved under the floating gate, and the output of the CCD settles to the signal voltage level. Then, the positive integration takes place. S3 and S5 open. S2 and S4 close connecting the output of the buffer to the non-inverting input of the integrator and the reference voltage to the inverting input of the integrator. Together, the two integration periods perform correlated double sampling (CDS) [9]. Finally, during the sampling period, only S5 closes making the last stage a voltage follower with a capacitor in the feedback loop, thus the output is the reference voltage plus the voltage across the capacitor.

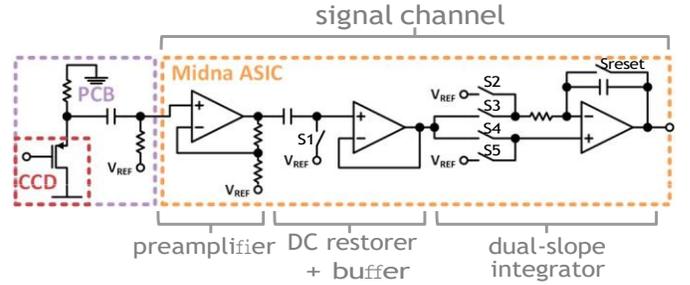

Fig. 1: Simplified schematic view of one of the four channels included in MIDNA. The channel includes a low-noise preamplifier, DC restore with buffer, and a dual-slope integrator.

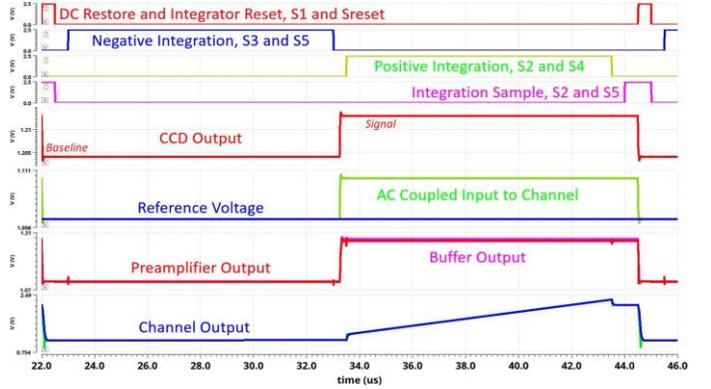

Fig. 2: Transient simulation showing the clocking and resulting signals in a MIDNA channel.

The specifications for the MIDNA channel are shown in Table I. The noise specifications are one-third of the expected output noise of the CCD including load resistor. The first large signal consideration is the baseline value of the CCD output. The process of DSI will cancel the DC value, but the large variation from pixel to pixel can lead to the integration signal going to the limits of the voltage range during the first integration slope. Even more, since the CCD signal is AC coupled, a voltage drift due to the CCD clocking may move the baseline voltage away from the reference voltage. Therefore, for all these reasons, a DC restorer block is required to ensure the baseline signal is close to the reference voltage at the integrator. Second, the integrator requires DC current to operate which cannot be provided through the DC restore capacitor. A buffer is necessary between the DC restorer and integrator to provide that current. Third, a DC restore capacitor of adequate size cannot be driven in sufficient time by the CCD output. A preamplifier is needed to provide the current to the DC restore. Additionally, chip area is saved by placing the most strict noise specifications on the preamplifier as opposed to the integrator.

Outside the chip, the connection to the CCD output requires a few PCB components for biasing and blocking of the high DC voltage. The resistive load for the CCD is chosen based on the CCD performance [18]. The AC coupling capacitor and resistor to $V_{REF}$ are the largest practical values for cryogenic operation, 100 nF and 100 kΩ. The reference voltage is provided by an external voltage regulator, LT3045. The regulator has 3 nV/$\sqrt{Hz}$ white noise and a corner frequency of 100 Hz in this use case.

TABLE I: Specifications for MIDNA

| SPECIFICATION | VALUE |
|---|---|
| Input Referred White Noise | 5 nV/$\sqrt{Hz}$ |
| Input Referred 1/f Noise | 1 µV/$\sqrt{Hz}$ at 1 Hz |
| Total Integration Time | 2 µs - 20 µs |
| Maximum Signal minus Baseline | 1.8 mV |

TABLE II: Input-referred Noise Spec. for MIDNA Blocks

| BLOCK | 1/F NOISE AT 1 HZ | WHITE NOISE |
|---|---|---|
| Preamplifier | 500 nV/$\sqrt{Hz}$ | 3 nV/$\sqrt{Hz}$ |
| Buffer | 1500 nV/$\sqrt{Hz}$ | 9 nV/$\sqrt{Hz}$ |
| Integrator | 1500 nV/$\sqrt{Hz}$ | 9 nV/$\sqrt{Hz}$ |

*B. Noise Considerations*

The MIDNA channel is designed for variable integration times, which can place the pass-band of the transfer function in either the white or 1/f noise region of the channel. As such, both the 1/f and white noise of the channel must be addressed. For the preamplifier and buffer, the noise transfer functions are scaled versions of Equation 1. The preamplifier function only scales by its gain, which is treated as constant across frequency. The buffer gain is the same as Equation 1. The DC restorer is treated as unity gain for simplicity. The noise specifications for the blocks are shown in Table II. The specifications on the buffer are relaxed considering the minimum preamplifier gain with some margin.

The input-referred noise from the integrator operational amplifier and resistor have a much different transfer function than the preamplifier and buffer because the gain is independent of the switch phases. The gain from the input of the operational amplifier to the output of the channel is below.

$$H_{n,int}(j\omega) = \frac{A}{j\omega}(1 - e^{2T_S j\omega}) \quad (4)$$

Importantly, the low frequency noise of the integrator is not filtered out. For standard DSI, this does not present a problem as the total contribution to the output is not dominant.

Multiplying the noise specifications for each block by the corresponding transfer functions in Equations 1 and 4, utilizing nominal gain settings, and integrating from 10 mHz to 10 MHz gives 110 µV$_{rms}$ of total noise at the output due to the channel. This is equivalent to the response from a 1 electron signal inside the CCD.

For the block designs, standard noise trade offs apply. These blocks use input pairs with size much greater than minimum to ensure the input referred noise of both types is low. The resistive feedback around the preamplifier uses a 200 Ω unit for a total resistance of 4 kΩ. Additionally, the resistor value of the integrator is only 75 kΩ. For the DC restorer, $kT/C$

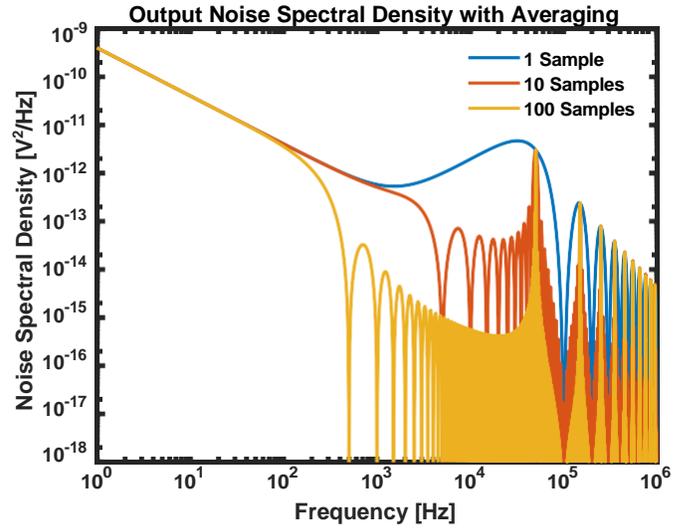

Fig. 3: The noise spectral density at the output of the channel. The noise around the frequencies of interest is greatly reduced with more samples averaged.

noise is not a first-order consideration because the process of DSI cancels it.

The noise characteristics of DSI specifically for non-destructive readout require additional considerations, especially for the integrator. As shown earlier, the traditional DSI band-pass filtering can be approximated into two pieces: the first-order high-pass filtering provided by the CDS action [19] and the first-order low-pass filtering of the integration function. As shown in Equation 4, the low frequency noise from the integrator is not filtered. Qualitatively, the CDS function is applied before the noise contribution of the operational amplifier and resistor inside the integrator.

In the DSI noise calculations so far, the contribution of the 1/f noise from the integrator was a negligible part of the total noise. However, non-destructive DSI readout uses averaging of samples of the output to combine all the reads. The DSI transfer function including sample averaging is

$$H_{DSI,avg}(j\omega) = \frac{A}{j\omega N}(1 - e^{-T_S j\omega})^2 \sum_{n=1}^{N} e^{-2nT_S} \quad (5)$$

where N is the number of consecutive samples used for the average. The transfer function to the output from the input of the integrator operational amplifier with averaging is

$$H_{n,int,avg}(j\omega) = \frac{A}{j\omega N}(1 - e^{2T_S j\omega}) \sum_{n=1}^{N} e^{-2nT_S} . \quad (6)$$

Averaging further lowers the bandwidth for the full chain around the frequencies of interest as plotted in Fig. 3. However, even with 100 averages, 1/f noise from the op amp inside the integrator remains as shown in Fig. 4. In the presence of 1/f noise from the integrator, the noise reduction from non-destructive readout saturates much like traditional DSI, Fig. 5.

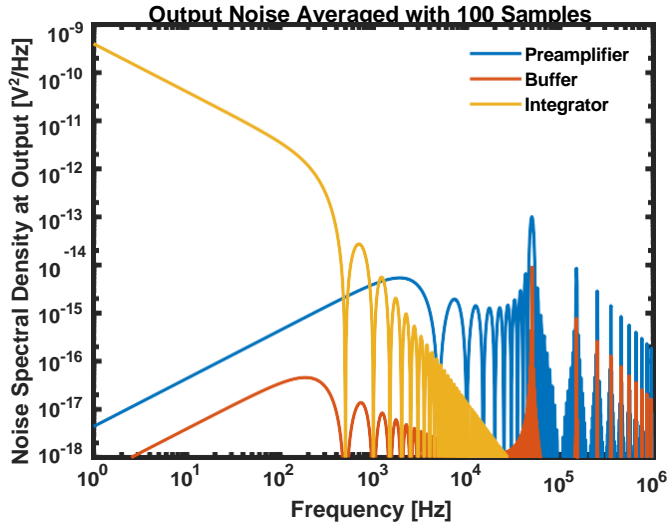

Fig. 4: The noise spectral density at the output of the channel after averaging 100 samples. The 1/f noise from the integrator is not significantly reduced.

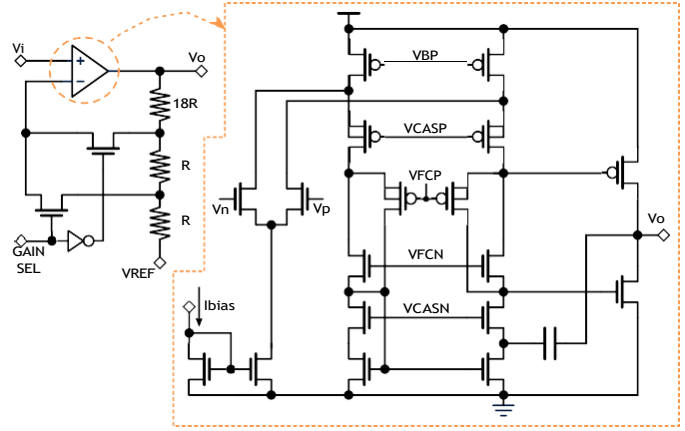

Fig. 6: Schematic of the low-noise preamplifier. It is implemented as a non-inverting amplifier with two gain selection switches in the feedback loop and referenced to VREF, which is the signal zero. The amplifier is a folded cascode amplifier with a push-pull class AB output stage that can drive the 100 pF DC restorer capacitance.

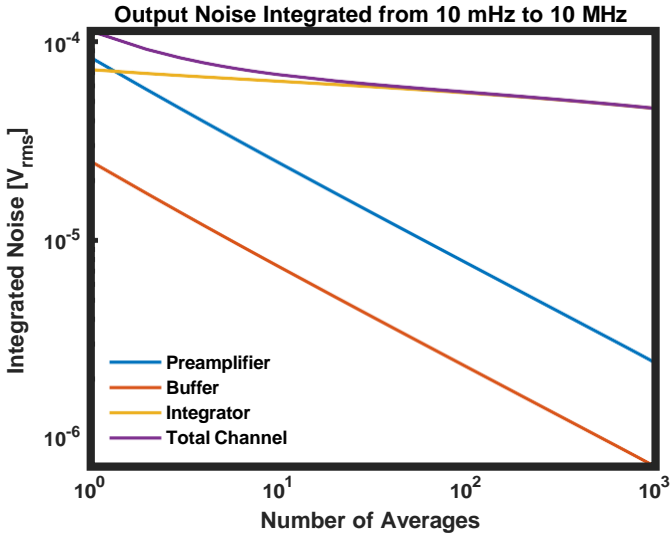

Fig. 5: The integrated noise at the output of the channel from 10 mHz to 10 MHz. The 1/f noise from the integrator dominates the total channel noise because the averaging does not eliminate it.

In the MIDNA ASIC, the inclusion of chopping inside the integrator's operational amplifier greatly reduces the 1/f noise from the integrator. Further discussion of this chopping is provided in the next section.

*C. Block Descriptions*

The preamplifier, shown schematically in Fig. 6, is largely responsible for the noise performance and settling the DC restorer block, which is a 100 pF capacitive load, in under 50 ns to allow fast averaging with the skipping technique. It is implemented as a folded cascode amplifier with large input devices and bias current. It takes 30% of the channel area and 76% of the power. To further enhance the settling, the output is implemented as a push-pull class AB stage that can deliver high currents. Finally, the block includes a resistive feedback network that defines the gain to be either a factor 10 or a factor 20, controlled by a single selection bit.

The DC restorer and the buffer are shown in Fig. 7. The DC restore action is carried out by connecting node(A to the reference voltage for a short period of time, turning on the switch S1, and then leaving that node in a high impedance state. The total series capacitance must be high enough to avoid a significant voltage drop during the whole DSI cycle due to leakage. Since the leakage depends on the signal voltage, a voltage drop that is signal dependent would ruin the DSI result. In MIDNA the series capacitor is implemented with Metal-Insulator-Metal (MIM) capacitors in parallel with Metal-Oxide-Metal (MOM) capacitors. The MOM capacitors use the fringe capacitance of metal levels from one up to six, whereas MIM capacitor are on top of MOMs, using metal levels seven and eight. The total capacitance is divided in two banks of 50 pF each that can be connected together for a total capacitance of 100 pF. A small signal drop appears due to the capacitor divider

The integrator's operational amplifier can be chopped by means of a dedicated input, POL. Fig. 8 shows how it is implemented on chip. The chopping allows the integration of the feedback amplifier's internal offset and low frequency noise in opposing directions during readout, greatly lowering the magnitude of the transfer function to the output of the channel. As a result, the output noise at low frequencies is reduced by over an order of magnitude as shown in Fig. 9.

Finally, charge injection is mitigated by using a large integration capacitance (20 pF), and its effect is also reduced by averaging either in the analog domain or in the digital

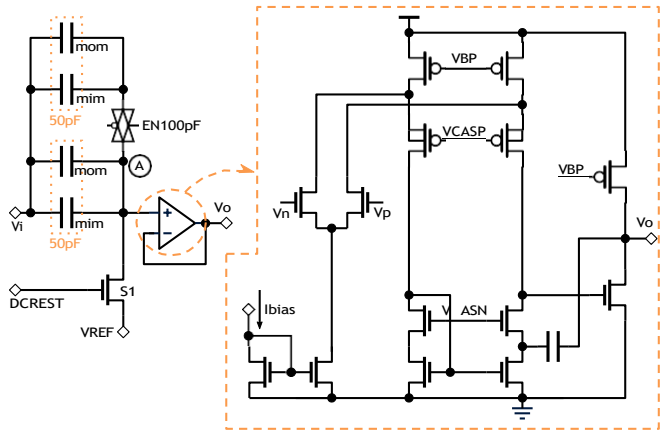

Fig. 7: DC restorer and buffer schematics. The DC restore includes Metal-Insulator-Metal (MIM) and Metal-Oxide-Metal (MOM) capacitors in parallel, increasing the total capacitance per unit of area. It is made in two separated banks of 50 pF which can be connected together. The buffer is a folded cascode amplifier similar to the preamplifier stage, but with a class A output.

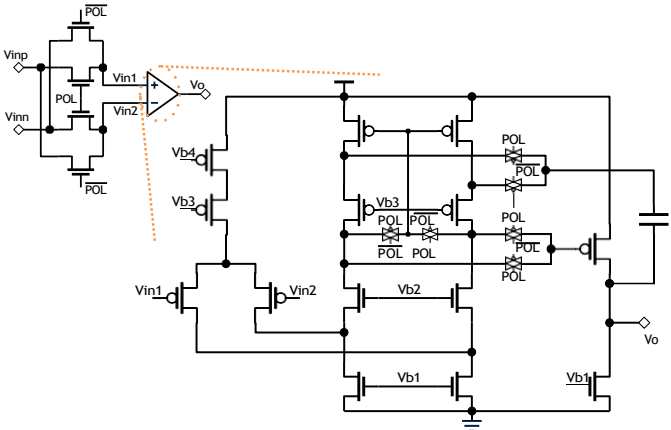

Fig. 8: Schematic of the integrator amplifier with chopping circuity that is key to enabling sub-electron noise performance. The chopping ensures that the low frequency noise of the integrator feedback amplifier does not saturate out the noise benefits of later low pass filtering.

domain, as it will be explained in the next section.

### D. Channel Control

MIDNA implements a flexible DSI readout system. First, the acquisition sequences and timings are not fixed. All timing phases can be controlled externally. Second, with MIDNA not only the standard differential averager signal processing can be carried out, i.e. sampling the output of the DSI and averaging the skipper samples in the digital domain, but MIDNA also allows the implementation of the analog pile-up technique demonstrated in [20], by which skipper samples are processed and averaged in the analog domain, reducing the number of analog to digital conversions and, therefore, the amount of data in a large scale dark matter experiment like OSCURA. Finally, MIDNA flexibility enables the bypassing of the DSI circuit to work in a *buffer* mode, in which the CCD output signal is only amplified without any other processing. This mode has been proven useful during the setting up and debug phase of an experiment.

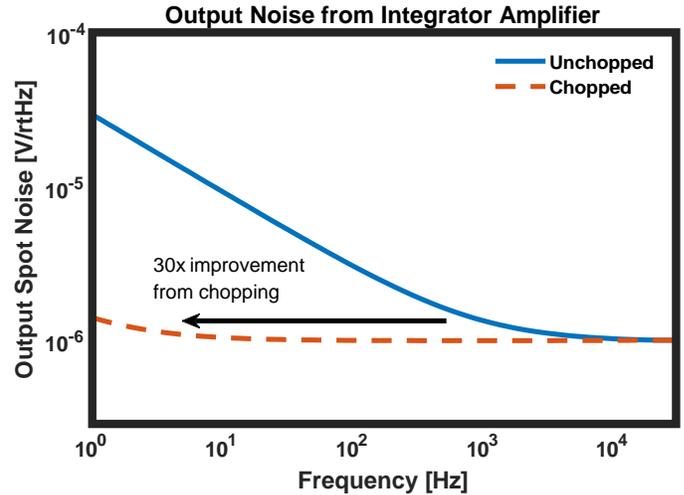

Fig. 9: Periodic noise simulation of the MIDNA chain with only the integrator noise present. The use of chopping decreases the low frequency noise contribution by more than an order of magnitude.

The flexibility of the control scheme allows for a variety of acquisition sequences. For comparison, a simplified standard DSI readout is depicted in Fig. 10a. At the end of the sequence a conversion from analog to digital occurs, which is represented in the figure as small dots over the MIDNA OUTPUT trace, and the sampled digital value is stored. The sequence is repeated for every skipper sample of the same pixel and the digital samples are averaged at the end of the acquisition on a post-processing step.

Fig. 10b shows an interesting variation of the standard sequence where there is no need of chopping of the integrator's operational amplifier and in which the output is sampled twice during the same skipper cycle. The first sample is the result of integrating only the baseline level, so if the integrator were ideal this sample would be at $V_{ref}$. However, since there are offsets and charge injection from the switches, there is a slight deviation from the ideal value. Then a reset pulse is applied and another sample is taken after a second pair of integrations equivalent to the standard sequence, which also includes offset and 1/f noise from the integrator. Thus, there are two samples per skipper cycle and both are stored by the controller. Finally, during the postprocessing step the difference between these samples is used as the skipper sample value, eliminating the non-ideal contributions by a digital CDS. The advantage of this *residual cancellation* sequence is that it eliminates the need of the amplifier chopping, but at the expense of twice the number of analog-to-digital acquisitions and a longer readout time.

The last sequence, depicted in Fig. 10c, is an implemen-

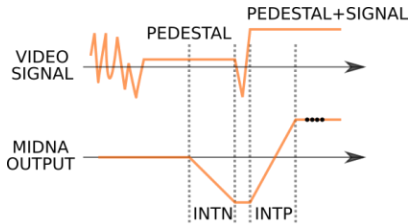

(a) Standard sequence for reading out one sample. The sequence includes one negative integration phase, one positive integration phase, and one sampling phase.

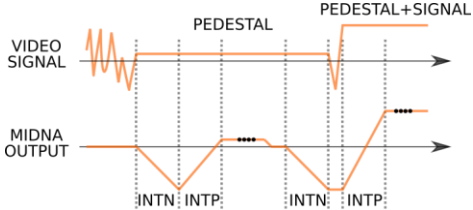

(b) Residual cancellation sequence. This sequence includes two negative integration phases, two positive integration phases, and two sampling phase. It does not require chopping for sub-electron noise performance but requires approximately double the time of the standard sequence.

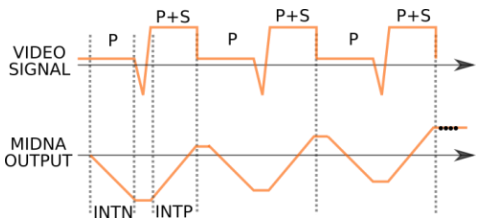

(c) Analog pile-up sequence. The example shows three skipper samples. This sequence can include many pairs of integration phases requiring only one sampling phase. With the use of chopping, sub-electron noise performance is attainable.

Fig. 10: A comparison of the three CCD video signal acquisition sequences that have been used in this work.

tation of the analog pile-up skipper-CCD readout technique described in [20]. The main differences with the standard readout are that the integrator is not reset between skipper samples and that only one analog to digital conversion is carried out after piling up $N$ skipper samples. This means that the output for one pixel is the addition of $N$ equal samples, which is equivalent to computing the average scaled by a factor $N$, and therefore the output noise is reduced by a factor of $1/\sqrt{N}$. So, the main advantage of this technique is that the averaging is done inside MIDNA by only adding skipper samples in the analog domain, and therefore, only one analog to digital conversion is needed. However, this comes at the expense of a reduced signal range. For $N$ skipper samples acquired, the signal range of the output gets reduced by a factor $N$, so while the theoretic signal range is 6000 e⁻ for one sample, piling up 1000 skipper samples lowers the range to 6 e⁻.

MIDNA is fabricated in a 65 nm low power CMOS process.

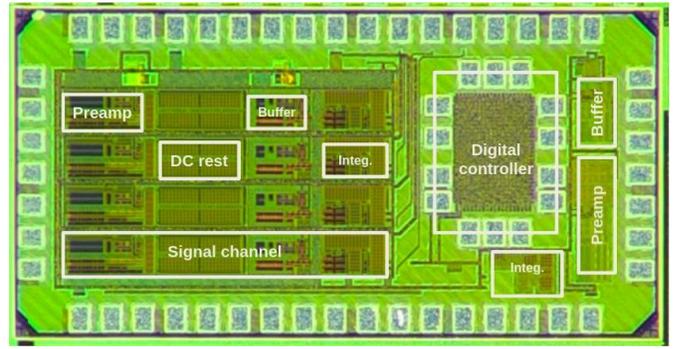

Fig. 11: Photograph of MIDNA. It is 2 mm by 1 mm. It includes four channels, a digital controller block and three standalone blocks for testing purposes. It was fabricated in a commercial 65 nm technology.

It was designed utilizing standard foundry models and models for 84 K [21] to ensure good performance at both room temperature and cryogenic temperatures. The availability of models for the cold case, which are already silicon proven, was the main reason for choosing this technology. All transistors were implemented with thick-gate devices in order to reduce leakage, which is further reduced by the low-temperature operation. A photograph of the fabricated chip is shown in Fig. 11. MIDNA includes four signal channels and is 2 mm by 1 mm. It also includes a digital synthesized logic that can generate the control signals for the analog part, and, for testing purposes, three standalone blocks: a preamplifier, an integrator and a buffer. The digital controller is separated from the analog part and its use is optional.

## III. EXPERIMENTAL RESULTS

Fig. 12 shows the experimental setup used to measure the ASIC at room temperature. It is comprised of a Low-Threshold Acquisition (LTA) board [2], a board containing the MIDNA ASIC and auxiliary electronics, i.e. voltage regulators, and a skipper-CCD inside the cold chamber. The skipper-CCD is cooled down to 140 K. A similar setup was used to test MIDNA cold, with the only difference that the MIDNA board was inside the cold chamber. The CCD output biasing resistor was 20 kΩ, the AC coupling capacitor 100 pF, and the MIDNA input bias resistor was 100 kΩ. All acquisitions shown in this work were made with the MIDNA gain set to minimum.

The LTA board generates the control signals for MIDNA and the CCD, and it also includes the analog-to-digital converters necessary to digitize the analog output of the ASIC. The LTA board has four input channels and each one was used to digitize the signal from one MIDNA channel. The LTA includes a differential preamplifier with an input range of ±1 V and a 18-bit 15 Msps ADC. The acquired data were formatted into a FITS image and transferred and stored in a computer with lossless compression.

Fig. 13 contains the results of a standard-sequence measurement of an AC-grounded channel with and without chopping of the integrator. Without mitigation of the 1/f noise contributed by the integrator feedback amplifier, the noise magnitude saturates versus number of samples as predicted in Fig. 5. The efficacy of chopping is evident. With the low frequency noise mitigated, the noise continues to be reduced by the number of averages, indicative of white noise.

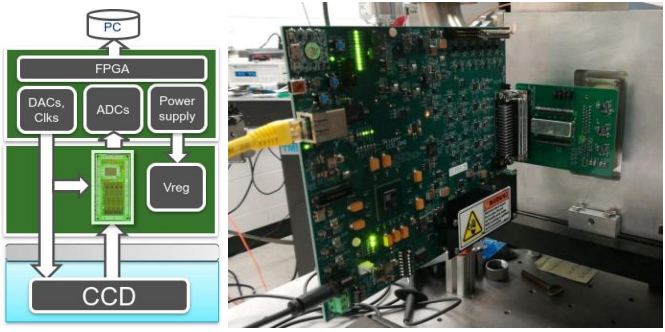

Fig. 12: Experimental setup for testing MIDNA at room temperature. It includes a skipper-CCD, MIDNA ASIC, and all the power supplies and control needed to operate them.

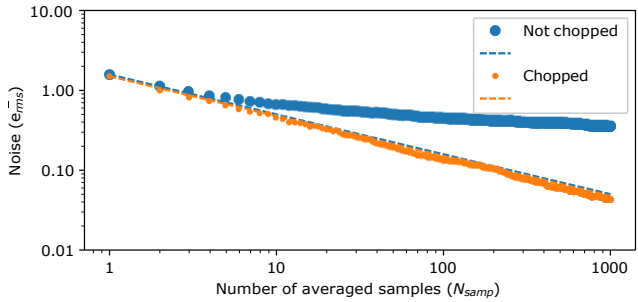

Fig. 13: Equivalent noise charge versus number of averaged samples $N_{samp}$ using the standard sequence. Chopping the integrator enables the noise to continue to fall with more averaged samples. The measurement is at at room temperature with the channel input AC grounded.

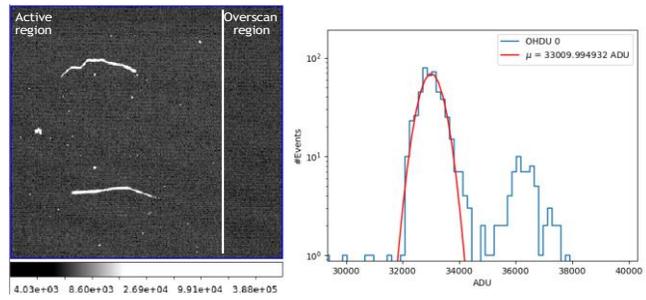

Fig. 14: Left: Example of acquired images showing the single photon ionization events and three ionization tracks. The *overscan* region is the result of reading out an imaginary extension of the real array and it only contains the noise generated by the CCD output stage. Right: Histogram of extracted single photon hits.

In order to fully verify the performance of the readout ASIC, a skipper-CCD is connected to the inputs of MIDNA, including the AC coupling passives. The skipper-CCD is designed by MicroSystem Labs of Lawrence Berkeley National Laboratory (LBNL) [1], [17]. It is a p-channel CCD fabricated on high-resistivity float-zone refined n-type silicon. The substrate is 675 µm thick with a resistivity of approximately 20 kΩ and, thus, it reaches full depletion with a bias voltage of 70 V. The sensor is divided in 1278×1058 square pixels with a pixel size of 15 µm. The sensitivity of this skipper-CCD is 1.8 µV/e⁻ in the setup conditions of the present work.

The gain of the whole acquisition system, from the CCD through MIDNA to the ADC, was measured by adding a low activity Iron-55 X-ray source in front of the CCD. Iron-55 decays via electron capture emitting 5.89 keV $K_\alpha$ photons with a probability of 16%, and 6.49 keV $K_\beta$ photons with a probability of 2.85% [22]. These photons interact with the silicon of the CCD by photoelectric effect producing on average 1570 electron-hole pairs and 1730 electron-hole pairs, respectively. These numbers were calculated taking into account 3.75 eV as the mean energy needed to generate an electron-hole pair reported in [23] for an skipper-CCD of similar characteristics working at 123 K.

Ten images were acquired and stored with the sensor exposed to the X-ray source during a one minute interval in-between readings. The standard sequence was used with an integration time of 13.3 µs for both positive and negative slopes of the DSI, and the total pixel acquisition time was 120 µs. This pixel time includes the intervals needed by the sensor to clear the spurious charge from the floating diffusion and also to move the charge packet under it with its associated long transients due to clock coupling. On a post-processing step, the mean of the overscan region was subtracted from the active part of the image and single photon ionization events were extracted and analyzed, computing a histogram of charge in analog-to-digital units (ADU). Fig. 14 shows an example image and the histogram of extracted hits. The $K_\alpha$ emission peak was fit with a Gaussian function in order to obtain the gain from its mean, and the readout RMS noise was calculated from the standard deviation of the samples in the overscan region. The gain result was 20.9 ADU/e⁻ with a single sample readout noise of 3.4 $e^-_{rms}$.

The highest contribution to the readout noise is the CCD. In order to verify this, the input of MIDNA was disconnected from the CCD and it was AC coupled to ground, close to the ASIC ground. A new acquisition was carried out without any other changes to the setup and the variance of the newly stored samples was computed. The result accounted for 35% of the total variance of the system, i.e. the $(3.4\,e^-_{rms})^2$, leaving the remaining 65% of the contribution coming from the sensor. This is in good agreement with the calculations and simulations performed during the ASIC design phase considering that the channel was designed to generate up to 30% of the CCD noise.

## A. Noise reduction by averaging skipper samples

The most important feature for the OSCURA experiment is to discriminate interactions that can generate charge packets of a few electrons and, therefore, a subelectron readout noise is necessary for the experiment. In order to reduce the readout noise, specifically the CCD output stage contribution, and to reach a subelectron noise, the averaging of skipper samples is needed. In this measurement subsection a skipper-CCD manufactured by another company was used, which is based on the same LBNL sensor design. These sensors were still in the development stage [6] during the time that this work was carried out, so the readout noise and dark current were not optimized. The skipper part of the sensor had a high charge transfer efficiency and, hence, it was possible to use it for the demonstration of noise reduction by averaging skipper samples.

The following paragraphs describe experimental details and results of three different acquisitions. The integration time was always set to 13.3 µs for both positive and negative integrations. An analysis of the data was carried out using custom python scripts to obtain the readout noise of the system. After computing the pixel histograms, a pair of single electron peaks was fitted with a function that is the addition of two Gaussian curves with different mean value. The difference in the mean represents the difference in the response to a charge of one electron, so it is directly associated to the gain of the whole system, expressed in ADU/e⁻. Additionally, the standard deviation obtained from the fit represents the readout root-mean-square noise of the system, which has ADC units. Then, the equivalent noise charge $\sigma_e$, in electrons RMS, was calculated by dividing the standard deviation by the gain.

*1) Standard sequence and digital averaging:* In the first experiment the acquisition was carried out using the standard sequence and with the integrator chopping enabled. One thousand samples per pixel were read out and stored. Then, on a post-processing step, the samples corresponding to the same pixel were averaged and histograms of pixel values were computed. Fig. 15 shows histograms for a number of averaged skipper samples of 200, 400, and 1000. Increasing the number of averaged samples reduces the readout noise, and so the charge discretization becomes evident. For the case of 1000 averaged samples, the resulting histogram shows a clear discretization of charge, and a single electron charge resolution is reached.

*2) Low temperature testing:* MIDNA was placed with the skipper-CCD inside the cold chamber. An temperature sensor located on the MIDNA board measured a temperature of 183 K. In this setup, the signal controlling chopping was not available, so the low frequency noise of the integrator was mitigated with the residual cancellation sequence instead. There were two analog to digital conversions per skipper sample and, therefore, twice the storage space and acquisition time was used.

*3) Analog pile-up technique:* Using a variation of the analog pile-up sequence, the skipper-CCD was readout with chopping and up to one thousand skipper samples per pixel were added into the integrator's capacitor. For this variation, there was an analog to digital conversion after every sample, storing all converted values. The advantage of storing every added skipper sample is that it is possible to analyze the noise reduction with an increasing number of skipper samples. Storing all samples is not needed in a real use case; only one analog to digital conversion is required at the end of the pile up sequence. By using only the last sample, the result is equivalent to the average of the skipper samples multiplied by a factor one thousand, so the noise is already reduced. Fig. 16 shows the readout noise versus the number of skipper samples, $N_{samp}$, that were piled up.

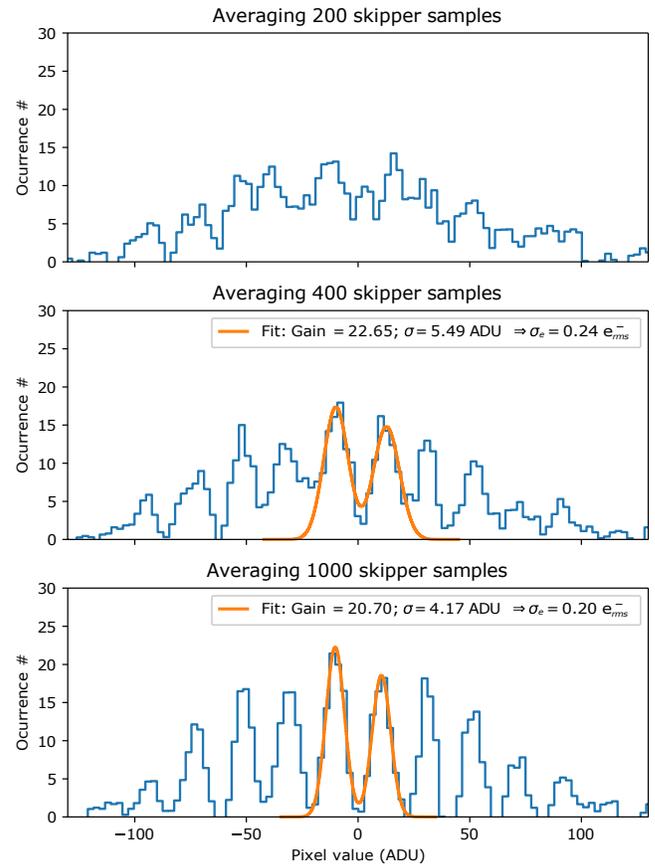

fig. 15: Pixel histograms for an increasing number of averaged samples (top to bottom). The standard acquisition sequence was used with MIDNA at room temperature. Single electron peaks were fitted with a function that is the addition of two Gaussian curves and the gain, in ADU/e⁻, and readout noise were extracted from the fit.

In plot of Fig. 16, the blue dot corresponds to a simple calculation of the pixel standard deviation without averaging, just using the first sample of each pixel, whereas the green dots correspond to the standard deviation of the Gaussian fit. The curve fitting routine provides an accurate result when the single electron peaks are clearly defined in the pixel histogram, and that happens when around 300 skipper samples or more have been averaged. The theoretical $1/\sqrt{N_{samp}}$ trend has also been plotted with a dashed line from $N_{samp}$ equal to 1, which means reading out an image with a single sample. There is a clear coincidence between the readout noise calculated from the single electron peaks fit and the theoretical noise reduction for the three tested sequences, meaning that averaging an increasing number of skipper samples reduces the noise by a factor $1/\sqrt{N_{samp}}$ as predicted. The calculated readout noise is $0.2\,e^-_{rms}$ after averaging one thousand skipper samples and is the same for the three acquisition sequences. Table III shows a summary of the experimental results. Note that in all cases, the CCD is the largest noise contributor, so the noise is not reduced with cooling of MIDNA.

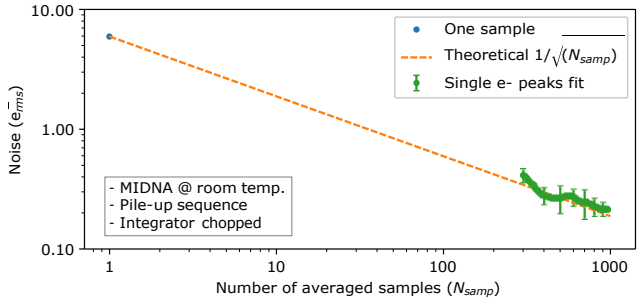

Fig. 16: Experimental demonstration of noise reduction versus the number of averaged samples ($N_{samp}$). The plot shows the case in which the analog pile-up technique, i.e. averaging in the analog domain, was carried out.

TABLE III: Noise reduction by averaging. Experimental results.

| EXPERIMENT | $N_{samp}$ 1 | $N_{samp}$ 1000 | Ratio |
|---|---|---|---|
| 1- Digital averaging | $5.2\,e^-$ | $0.194\,e^-$ | 0.037 |
| 2- MIDNA cooled down | $6.0\,e^-$ | $0.206\,e^-$ | 0.034 |
| 3- Analog averaging | $5.9\,e^-$ | $0.196\,e^-$ | 0.033 |

A comparison with other CCD readout ASICs has been made on Table IV. The main difference is the inclusion of the DSI circuit in MIDNA, which reduces the input referred readout noise voltage more than a factor ten with respect to others and, therefore, allows an ultra-low noise acquisition with a low sensitivity CCD. On the contrary, other ASICs make use of different methods. For example, the MCRC ASIC [24] does the CDS by amplifying the difference between pedestal and signal leves, and the MND0x ASIC series [25], [26] include a customized delta-sigma modulator with a response similar to the DSI. The advantage of MIDNA is that it is also capable of the pile-up function, which is specifically intended for non-destructive readout of skipper-CCDs. The ASIC has been tested down to 180 K and the power consumption per channel is only 4.2 mW, which is a key point for the OSCURA project since it will be inside the cold vessel close to the sensors. Regarding readout speed, in the tests that were carried out the speed was always limited by the sensor. Although the integration time was set to 13.3 µs, most of the time was spent waiting for the CCD output voltage to settle, which greatly reduces the readout speed, so the total time per sample was 120 µs. However, MIDNA is capable of 1 µs integration time with negligible deadtime, and so it can run up to 500 kpixel/s.

## IV. CONCLUSIONS

MIDNA is a low-noise integrated circuit that implements a skipper-CCD readout system based on a dual-slope integrator architecture and that can operate under cryogenic temperatures next to the sensor. It is intended to replace the current discrete electronics used for reading out skipper-CCDs and enable experiments that use these sensors on a large scale. For the particular case of the OSCURA experiment, the low noise performance in a scalable architecture will allow for the build up of the 24000 skipper-CCDs (28 Gpix) dark matter detector.

The design of the MIDNA channel allows the implementation of different readout schemes only by programming the control signals of the switches. Three acquisition sequences have been presented here: the standard DSI sequence; the residual cancellation, which allows the reduction of noise by a digital CDS of the analog integration; and the implementation of the analog pile-up technique that allows the reduction of the amount of analog to digital conversions by averaging in the analog domain, at the expense of a reduced signal range.

The three readout sequences proved useful to reduce the noise as an increasingly large number of skipper samples is averaged. It has been shown that the noise reduction follows the $1/\sqrt{N}$ scaling up to N equal to 1000 skipper samples, reaching the single electron charge resolution with a readout noise of $0.2\,e^-_{rms}$.

## REFERENCES

[1] J. Tiffenberg et al., "Single-electron and single-photon sensitivity with a silicon skipper CCD," *Physical Review Letters*, vol. 119, no. 13, p. 131802, 2017.

[2] G. I. Cancelo, C. Chavez, F. Chierchie, J. Estrada, G. Fernandez-Moroni, E. E. Paolini, M. S. Haro, A. Soto, L. Stefanazzi, J. Tiffenberg, K. Treptow, N. Wilcer, and T. J. Zmuda, "Low threshold acquisition controller for Skipper charge-coupled devices," *Journal of Astronomical Telescopes, Instruments, and Systems*, vol. 7, no. 1, pp. 1 – 19, 2021. [Online]. Available: https://doi.org/10.1117/1.JATIS.7.1.015001

[3] L. Barak, I. M. Bloch, M. Cababie, G. Cancelo, L. Chaplinsky, F. Chierchie, M. Crisler, A. Drlica-Wagner, R. Essig, J. Estrada *et al.*, "SENSEI: Direct-detection results on sub-gev dark matter from a new skipper CCD," *Physical Review Letters*, vol. 125, no. 17, p. 171802, 2020.

[4] A. Aguilar-Arevalo, D. Amidei, D. Baxter, G. Cancelo, B. C. Vergara, A. Chavarria, E. Darragh-Ford, J. de Mello Neto, J. D'Olivo, J. Estrada *et al.*, "Constraints on light dark matter particles interacting with electrons from DAMIC at SNOLAB," *Physical review letters*, vol. 123, no. 18, p. 181802, 2019.

[5] J. D'Olivo, C. Bonifazi, D. Rodrigues, and G. F. Moroni, "vIOLETA: neutrino interaction observation with a low energy threshold array," in *XXIX Int. Conf. Neutrino Phys*, 2020.

[6] A. Aguilar-Arevalo, F. A. Bessia, N. Avalos, D. Baxter, X. Bertou, C. Bonifazi, A. Botti, M. Cababie, G. Cancelo, B. A. Cervantes-Vergara, N. Castello-Mor, A. Chavarria, C. R. Chavez, F. Chierchie, J. M. De Egea, J. C. D'Olivo, C. E. Dreyer, A. Drlica-Wagner, R. Essig, J. Estrada, E. Estrada, E. Etzion, G. Fernandez-Moroni, M. Fernandez-Serra, S. Holland, A. L. Barreda, A. Lathrop, J. Lipovetzky, B. Loer, E. M. Villalpando, J. Molina, S. Perez, P. Privitera, D. Rodrigues,

TABLE IV: MIDNA Performance Results and Comparison

| Parameter | Observations | MIDNA | MCRC[24] | MND0x[25], [26] | Units |
|---|---|---|---|---|---|
| Technology node | | 65 | 350 | 350 | nm |
| Supply voltage | | 2.5 | 3.3 | 3.3 | V |
| Method | | DSI, Pile-up | Diff. amp. | Custom delta-sigma | — |
| Integration time | | 1 to 20 | — | — | µs |
| Input range V | Int. time: 10 µs | 15 | 320 | 40 | mV |
| CCD sensitivity | Used for testing | 1.8 | 35 | — | µV/e$^-$ |
| Input range e$^-$ | Int. time: 10 µs | 8300 | 9140 | — | e$^-$ |
| Input referred noise | Int. time: 13.3 µs | 3.6 | 57 | 50 | µV$_{rms}$ |
| Max. readout speed | Int. time: 1 µs | 500 | 5000 | 1000 | kpixel/s |
| Temperature range | | 180 to 297 | Not reported | 297 | K |
| Power per channel | | 4.2 | 30.8 | 38 | mW |

R. Saldanha, D. S. Cruz, A. Singal, N. Saffold, L. Stefanazzi, M. Sofo-Haro, J. Tiffenberg, C. Torres, S. Uemura, and R. Vilar, "The oscura experiment," 2022. [Online]. Available: https://arxiv.org/abs/2202.10518

[7] J. Estrada, "Observatory of Skipper CCDs Unveiling Recoiling Atoms (OSCURA)," Available: https://astro.fnal.gov/science/dark-matter/oscura/, Accessed: 2022-07-01.

[8] M. White, D. Lampe, F. Blaha, and I. Mack, "Characterization of surface channel ccd image arrays at low light levels," IEEE Journal of Solid-State Circuits, vol. 9, no. 1, pp. 1–12, 1974.

[9] J. Janesick, Scientific charge-coupled devices. SPIE press, 2001, vol. 83.

[10] M. Kłosowski, W. Jendernalik, J. Jakusz, G. Blakiewicz, and S. Szczepański, "A cmos pixel with embedded adc, digital cds and gain correction capability for massively parallel imaging array," IEEE Transactions on Circuits and Systems I: Regular Papers, vol. 64, no. 1, pp. 38–49, 2017. [Online]. Available: https://doi.org/10.1109/TCSI.2016.2610524

[11] J. Cheon and G. Han, "Noise analysis and simulation method for a single-slope adc with cds in a cmos image sensor," IEEE Transactions on Circuits and Systems I: Regular Papers, vol. 55, no. 10, pp. 2980–2987, 2008. [Online]. Available: https://doi.org/10.1109/TCSI.2008.923434

[12] D. G. Chen, F. Tang, M.-K. Law, X. Zhong, and A. Bermak, "A 64 fj/step 9-bit sar adc array with forward error correction and mixed-signal cds for cmos image sensors," IEEE Transactions on Circuits and Systems I: Regular Papers, vol. 61, no. 11, pp. 3085–3093, 2014. [Online]. Available: https://doi.org/10.1109/TCSI.2014.2334852

[13] K. Park, S. Yeom, and S. Y. Kim, "Ultra-low power cmos image sensor with two-step logical shift algorithm-based correlated double sampling scheme," IEEE Transactions on Circuits and Systems I: Regular Papers, vol. 67, no. 11, pp. 3718–3727, 2020. [Online]. Available: https://doi.org/10.1109/TCSI.2020.3012980

[14] D. J. Hegyi and A. Burrows, "Optimal sampling of charge-coupled devices," The Astronomical Journal, vol. 85, pp. 1421–1424, 1980.

[15] G. R. Hopkinson and D. H. Lumb, "Noise reduction techniques for CCD image sensors," Journal of Physics E: Scientific Instruments, vol. 15, no. 11, pp. 1214–1222, nov 1982. [Online]. Available: https://doi.org/10.1088/0022-3735/15/11/020

[16] G. Fernández Moroni, J. Estrada, G. Cancelo, S. E. Holland, E. E. Paolini, and H. T. Diehl, "Sub-electron readout noise in a skipper CCD fabricated on high resistivity silicon," Experimental Astronomy, vol. 34, no. 1, pp. 43–64, 2012. [Online]. Available: https://doi.org/10.1007/s10686-012-9298-x

[17] S. E. Holland, D. E. Groom, N. P. Palaio, R. J. Stover, and M. Wei, "Fully depleted, back-illuminated charge-coupled devices fabricated on high-resistivity silicon," IEEE Transactions on Electron Devices, vol. 50, no. 1, pp. 225–238, 2003.

[18] S. Haque, F. Dion, R. Frost, R. Groulx, S. E. Holland, A. Karcher, W. F. Kolbe, N. A. Roe, G. Wang, and Y. Yu, "Design of low-noise output amplifiers for P-channel charge-coupled devices fabricated on high-resistivity silicon," in Sensors, Cameras, and Systems for Industrial and Scientific Applications XIII, R. Widenhorn, V. Nguyen, and A. Dupret, Eds., vol. 8298, International Society for Optics and Photonics. SPIE, 2012, p. 82980X. [Online]. Available: https://doi.org/10.1117/12.905460

[19] R. Kansy, "Response of a correlated double sampling circuit to 1/f noise [generated in ccd arrays]," IEEE Journal of Solid-State Circuits, vol. 15, no. 3, pp. 373–375, 1980.

[20] M. S. Haro, C. Chavez, J. Lipovetzky, F. A. Bessia, G. Cancelo, F. Chierchie, J. Estrada, G. F. Moroni, L. Stefanazzi, J. Tiffenberg, and S. Uemura, "Analog pile-up circuit technique using a single capacitor for the readout of skipper-CCD detectors," Journal of Instrumentation, vol. 16, no. 11, p. P11012, nov 2021. [Online]. Available: https://doi.org/10.1088/1748-0221/16/11/p11012

[21] D. Braga, S. Li, and F. Fahim, "Cryogenic electronics development for high-energy physics: An overview of design considerations, benefits, and unique challenges," IEEE Solid-State Circuits Magazine, vol. 13, no. 2, pp. 36–45, 2021.

[22] A. Thompson, D. Attwood, E. Gullikson, M. Howells, K. Kim, J. Kirz, J. Kortright, I. Lindau, Y. Liu, P. Pianetta, A. Robinson, J. Scofield, J. Underwood, and G. Williams, X-ray data booklet. Lawrence Berkeley National Laboratory, University of California Berkeley, CA, 2001, vol. 8, no. 4.

[23] D. Rodrigues, K. Andersson, M. Cababie, A. Donadon, A. Botti, G. Cancelo, J. Estrada, G. Fernandez-Moroni, R. Piegaia, M. Senger, M. S. Haro, L. Stefanazzi, J. Tiffenberg, and S. Uemura, "Absolute measurement of the Fano factor using a skipper-CCD," Nuclear Instruments and Methods in Physics Research Section A: Accelerators, Spectrometers, Detectors and Associated Equipment, vol. 1010, p. 165511, 2021. [Online]. Available: https://www.sciencedirect.com/science/article/pii/S0168900221004964

[24] P. Orel, S. Herrmann, T. Chattopadhyay, G. R. Morris, S. W. Allen, G. Y. Prigozhin, R. Foster, A. Malonis, M. W. Bautz, M. J. Cooper, and K. Donlon, "X-ray speed reading with the MCRC: a low noise CCD readout ASIC enabling readout speeds of 5 Mpixel/s/channel," in X-Ray, Optical, and Infrared Detectors for Astronomy X, A. D. Holland and J. Beletic, Eds., vol. 12191, International Society for Optics and Photonics. SPIE, 2022, p. 1219124. [Online]. Available: https://doi.org/10.1117/12.2629049

[25] H. Nakajima, D. Matsuura, T. Idehara, N. Anabuki, H. Tsunemi, J. P. Doty, H. Ikeda, H. Katayama, H. Kitamura, and Y. Uchihori, "Development of the analog asic for multi-channel readout x-ray ccd camera," Nuclear Instruments and Methods in Physics Research Section A: Accelerators, Spectrometers, Detectors and Associated Equipment, vol. 632, no. 1, pp. 128–132, 2011. [Online]. Available: https://www.sciencedirect.com/science/article/pii/S0168900210029761

[26] H. Nakajima, S. nosuke Hirose, R. Imatani, R. Nagino, N. Anabuki, K. Hayashida, H. Tsunemi, J. P. Doty, H. Ikeda, H. Kitamura, and Y. Uchihori, "Development of low-noise high-speed analog asic for x-ray ccd cameras and wide-band x-ray imaging sensors," Nuclear Instruments and Methods in Physics Research Section A: Accelerators, Spectrometers, Detectors and Associated Equipment, vol. 831, pp. 283–287, 2016, proceedings of the 10th International "Hiroshima" Symposium on the Development and Application of Semiconductor Tracking Detectors. [Online]. Available: https://www.sciencedirect.com/science/article/pii/S0168900216302121